\documentclass[ reprint, amsmath,amssymb, aps,pra]{revtex4-1}
\usepackage{tikz}
\usepackage{amscd}
\usepackage{amsthm}
\usepackage{graphicx}
\usepackage{mathdots}
\usepackage{epstopdf}
\usepackage{enumerate}
\usepackage{changepage}

\usepackage[
colorlinks,
linkcolor = blue,
citecolor = blue,
urlcolor = blue]{hyperref}
\def \qed {\hfill \vrule height7pt width 7pt depth 0pt}

\newtheorem{theorem}{Theorem}
\newtheorem{corollary}{Corollary}
\newtheorem{lemma}{Lemma}
\newtheorem{example}{Example}
\newtheorem{definition}{Definition}
\usepackage{dcolumn}
\usepackage{bm}

\begin{document}

\preprint{APS/123-QED}

\title{Alternative method for deriving nonlocal  multipartite  product states}

\author{Mao-Sheng Li}
 \affiliation{
 Department of Mathematical Sciences, Tsinghua University, Beijing100084, China}
\author{Yan-Ling Wang}%
 \email{wangylmath@yahoo.com}
\affiliation{
 School of Computer Science and Network Security, Dongguan University of Technology, Dongguan 523808, China
}

\begin{abstract}
	  We can only perform a finite rounds of  measurements  in  protocols  with  local operations and classical communication (LOCC).  In this paper,    we propose a set of product states, which require infinite rounds of measurements in order to distinguish this given set of states perfectly by LOCC.  Therefore, we can conclude that these sets of states are locally indistinguishable. More accurately,
given any multipartite LOCC indistinguishable set where every
local system cannot start with a nontrivial measurement, then after
appending these states with arbitrarily choosing two nonorthogonal
states, we  obtain another LOCC indistinguishable set.
		 It can be seen that   some parties can perform some nontrivial measurements. Hence, these sets are quite different from those constructed before.  This result broadens the knowledge of nonlocality without entanglement to a certain extent.

	\begin{description}
\item[PACS numbers] 03.67.Hk,03.65.Ud
\end{description}
 \end{abstract}                            
\maketitle

	\section{Introduction}
In quantum information theory,  quantum information is always hidden in the quantum states. In particular, classical information can also be encoded within a set of finitely chosen quantum states. The decoding process can be seen as a protocol that could distinguish the encoded quantum states. It is well known that a given set of quantum states can be perfectly identified if and only if the states of the given set are mutually orthogonal \cite{nils}. However, we often encounter compound quantum systems, and the physical conditions always restrict our ability such that we are only allowed to do local operations and classical communication (LOCC).  In fact, there exist many global operators that cannot be implemented by LOCC. Hence, it is fundamentally of interest to study the following problem: given a set of mutually orthogonal states in compound quantum systems, can these given states be identified only using LOCC? If so, we call them locally distinguishable or distinguishable by LOCC. If not, we call them locally indistinguishable or indistinguishable by LOCC.  These states are also known as a nonlocal set in the latter case. The local distinguishability has been practically applied in quantum cryptography primitives such as secret sharing and data hiding \cite{DiVincenzo02,Markham08}.
	
	Maximally entangled states and product states have been mostly studied. References \cite{Gho01,Wal00,Wal02,Fan04,Nathanson05,Cohen07,Bandyopadhyay11,Li15,Fan07,Yu12,Cos13,Yu115} provide an incomplete list of the results about the local distinguishability of maximally entangled states. At the same time, many people have considered the local distinguishability of product states \cite{Ben99,Ran04,Hor03,Ben99b,DiVincenzo03,Zhang14,Zhang15,Zhang16,Xu16b,Xu16m,Zhang16b,Wang15,Wang17,Feng09,Yang13,Zhang17,Zhangj17,Halder18}.
	
	The case of product states was first considered by Bennett.\emph{ et al}., who presented nine LOCC indistinguishable product states in $\mathbb{C}^3\otimes\mathbb{C}^3$ \cite{Ben99}. Since then, the local indistinguishability
	of orthogonal product states in bipartite systems has attracted much attention \cite{Feng09,Zhang14,Zhang15,Wang15,Zhang16,Xu16b,Zhang16b}. More recently, the study of multipartite quantum systems has attracted increased attention \cite{Xu16m,Wang17,Zhang17,Halder18}. More precisely, Xu \emph{et al.} presented a small set with $2n$ locally indistinguishable multipartite orthogonal
	product states in $\mathbb{C}^{d_1}\otimes\mathbb{C}^{d_2}\otimes\cdots\otimes\mathbb{C}^{d_n}$  for $n\geq 3$. Subsequently, Wang \emph{et al.} gave an explicit construction of locally indistinguishable multipartite orthogonal product states
	for multipartite quantum systems using a set of locally
	indistinguishable bipartite orthogonal product states. Zhang \emph{et al.} provided general methods to construct locally
	indistinguishable orthogonal product states for multipartite quantum systems using a set of locally indistinguishable bipartite orthogonal product states with the following additional condition: two partites cannot start a nontrivial measurement. More recently, S. Halder presented several sets that no states can be eliminated from the basis by performing orthogonality preserving measurements \cite{Halder18}. Another direction of related research is to study entanglement as a resource to distinguish quantum states of locally indistinguishable states \cite{Cohen08,Bandyopadhyay16,Zhang16E}.
	
	Note that for almost all these results, except the unextendible product basis(UPB) \cite{Ben99b,DiVincenzo03,Feng06,J14,CJ15}, their construction is based on useful techniques developed by Walgate and Hardy \cite{Wal00}. Hence, it is fundamentally of interest to find some other methods that would lead to more abundant constructions of nonlocal product states.

	The rest of the article is organized as follows. In Sec. \ref{sec}, we give some necessary definitions and some  important concepts. In Sec. \ref{thi}, we present our main results.  We start with a simple example to show the main idea of our method. Then we show a much more general statement.  For some  given   set of mutually orthogonal product states which is LOCC indistinguishable because all the local systems can not start with a nontrivial measurement, we can extend it  by another partite whose local state is selected from a set with two nonorthogonal states.   Although with one more partite joined in, the set is still LOCC indistinguishable.  After that, we generalize this with more parties joined in and show the   target set is still LOCC indistinguishable. At last, we give an example to show that our theorems can not be ``simply generalized" to a set of three or more   nonorthogonal states.  Finally, we draw a conclusion and present some interesting problems in     section \ref{for}.
\section{Preliminaries}\label{sec}
To  perfectly distinguish  a set of states, these states must be orthogonal to each other. Based on this fact, we know that only orthogonality-preserving measurements are allowed   to  perfectly  distinguish   a set of orthogonal states by LOCC.
	Most locally indistinguishable product states are based on the principle that  there must exist some partite who can be the first to perform a nontrivial measurement. Here, we first give the definition of trivial measurement.

	\begin{definition}\label{trivial}
		A measurement $\{M_i\}_{i=1}^n$ corresponding to a system of level $d$ is called a trivial measurement if
		$$ \sum_{i=1}^nM_i^\dagger M_i=I_d  \text{ and  }\ \ \ M_i^\dagger M_i\propto I_d $$
		for any $1 \leq i\leq n$.
	\end{definition}
	
	In order to perfectly distinguish a set of orthogonal states in multipartite quantum systems, any measurement should preserve the orthogonality relations of given states. The following six states in $ \mathbb{C}^2\otimes \mathbb{C}^2\otimes \mathbb{C}^2 $ can be proved to be LOCC indistinguishable, since no partite could start with a nontrivial measurement in order to preserve the orthogonality relations of these given states (See Refs. \cite{Feng09,Xu16m}).
	\begin{equation}\label{states}
	\begin{array}{c}
	|\psi_1\rangle=|0\rangle|1\rangle|0+1\rangle, \\
	|\psi_2\rangle=|0\rangle|1\rangle|0-1\rangle,\\
	|\psi_3\rangle=|0+1\rangle|0\rangle|1\rangle,\\
	|\psi_4\rangle=|0-1\rangle|0\rangle|1\rangle,\\
	|\psi_5\rangle=|1\rangle|0+1\rangle|0\rangle,\\
	|\psi_6\rangle=|1\rangle|0-1\rangle|0\rangle.
	\end{array}
	\end{equation}

	\begin{definition}\label{completable}
		Let $\mathcal{H}$ denote  an $m$ partite quantum system $ \otimes_{i=1}^m\mathcal{H}_i$.  Suppose $\mathcal{S}$ is a set of mutually orthogonal product states in $\mathcal{H}$. We say $\mathcal{S}$ is completable in $\mathcal{H}$ if its complementary subspace $\mathcal{S}^\perp$  is spanned by a set of orthogonal product states. Equivalently, there exists a product basis $\mathcal{B}$ of $\mathcal{H}$ such that  $\mathcal{S}\subseteq \mathcal{B}$.
	\end{definition}
	
	\section{method for deriving Nonlocal  multipartite  product states}\label{thi}
	
	It is well known that  a set $\mathcal{S}$ of  quantum states  can be distinguished by global operations if and only if   the states in $\mathcal{S}$ are pairwise orthogonal to each other.
	The following lemma illustrates the most important reason why two nonorthogonal quantum states cannot be distinguished.  This lemma will be used repeatedly in this article. So we also give a detail proof of it. But for the sake of readability of the article we leave it  in APPENDIX A.
	\begin{lemma}\label{lemm}
		Let $|\alpha\rangle, |\beta\rangle$ be two nonorthogonal states in $\mathbb{C}^{d}$. For any  measurement $\mathcal{M}=\{ M_i\}_{i=1}^n$, there exists some  $i$ such that the following three terms are nonzero:
		$$
		\langle \alpha|  M_i^\dagger M_i|\alpha\rangle,\langle \alpha|  M_i^\dagger M_i|\beta\rangle,\langle \beta|M_i^\dagger   M_i|\beta\rangle.$$
That is, the post-measurement states of the outcome $i$ are still nonorthogonal states.
	\end{lemma}

Now we begin to discuss the main issues we will consider.	
	Notice that any five of the six states in equation (\ref{states}) are LOCC distinguishable. Therefore, it is easy to see that the following six states
	\begin{equation}\label{statesset}|\psi_1\rangle|\alpha_1\rangle,|\psi_2\rangle|\alpha_2\rangle,|\psi_3\rangle|\alpha_3\rangle,|\psi_4\rangle|\alpha_4\rangle,|\psi_5\rangle|\alpha_5\rangle,
	|\psi_6\rangle|\alpha_6\rangle
	\end{equation}
	are LOCC distinguishable if  two of the six states $ |\alpha_1\rangle,$ $|\alpha_2\rangle,$ $|\alpha_3\rangle,$ $|\alpha_4\rangle,$ $|\alpha_5\rangle,$ $ |\alpha_6\rangle $   are orthogonal to each other. It is interesting to consider the following case: Suppose $ |\alpha_1\rangle,$ $|\alpha_2\rangle,$ $|\alpha_3\rangle,$ $|\alpha_4\rangle,$ $|\alpha_5\rangle,$ $ |\alpha_6\rangle $  are not orthogonal to each other, then what result about the  local distinguishability of states  in (\ref{statesset}) can we obtain?  Here we start with a simple example.

\begin{example}\label{exam1} With the same notation in equation (\ref{states}),
		and let $|\alpha\rangle,|\beta\rangle $ be any two nonorthogonal states in $\mathbb{C}^2.$
		Then the set of states $$\mathcal{S}:=\{|\psi_1\rangle|\alpha\rangle,|\psi_2\rangle|\alpha\rangle,|\psi_3\rangle|\alpha\rangle,|\psi_4\rangle|\beta\rangle,|\psi_5\rangle|\beta\rangle,
		|\psi_6\rangle|\beta\rangle\}$$
		in $ \mathbb{C}^2\otimes \mathbb{C}^2\otimes \mathbb{C}^2\otimes \mathbb{C}^2$ is LOCC indistinguishable .
	\end{example}	
	\noindent\emph{ Proof}: Denote   $$ |\alpha_1\rangle=|\alpha_2\rangle=|\alpha_3\rangle=|\alpha\rangle, \ |\alpha_4\rangle=|\alpha_5\rangle=|\alpha_6\rangle=|\beta\rangle. $$
Since the last partite  of states in $\mathcal{S}$ are not orthogonal to each other, we can show that  all the first three  parties could only start with a  trivial measurement in order to preserve the orthogonality relations of   states in $\mathcal{S}$. For instance, if the first partite goes first in the distinguishing protocol.  Let $M^\dagger M$ be  one of the POVM matrix performed by the first partite.  The orthogonal relations
{ $$ \langle \alpha_i|\langle \psi_i| M^\dagger M\otimes I_2 \otimes I_2\otimes I_2|\psi_j\rangle|\alpha_j\rangle=0,   \text { for all } i\neq j$$}
are just equivalent with $$   \langle \psi_i| M^\dagger M\otimes I_2 \otimes I_2 |\psi_j\rangle=0, \text { for all } i\neq j.$$
  However, using the latter relations, we can deduce that $M^\dagger M\propto I_2$, i.e., the first partite could only start with trivial measurement.

 Hence,  the fourth   partite should be  the first one to take a nontrivial measurement $\mathcal{M}=\{M_i\}_{i=1}^n$.
	By lemma \ref{lemm}, there exists at least one $i$ such that
	$|\alpha'\rangle=M_i|\alpha\rangle, |\beta'\rangle= M_i|\beta\rangle$ are not orthogonal to each other. Normalizing the two states $|\alpha'\rangle, |\beta'\rangle$ to be $|\widehat{\alpha}'\rangle, |\widehat{\beta}'\rangle$, we redenote them to be $|\alpha'\rangle, |\beta'\rangle$.
	After the first round of measurement with the outcome $\text{``}i \text{ ''}$,  the post-measurement states are just
{\small	$$\mathcal{S}'=\{|\psi_1\rangle|\alpha'\rangle,|\psi_2\rangle|\alpha'\rangle,|\psi_3\rangle|\alpha'\rangle,|\psi_4\rangle|\beta'\rangle,|\psi_5\rangle|\beta'\rangle,
		|\psi_6\rangle|\beta'\rangle\}.$$
 }
 In order to perfectly distinguish the states $\mathcal{S}$, we have to perfectly distinguish the states $ \mathcal{S}'$. When  compared the set $\mathcal{S}$ with $\mathcal{S}'$, we realize that they share  the same property :   the first three local systems cannot start with   a nontrivial measurement, and the local states of the fourth system are just chosen from a set with two nonorthogonal states.

 With a  similar argument for the set $ \mathcal{S}'$,   in the second round of measurement, there exists an outcome $\text{``}j \text{ ''}$ of the fourth system measurement such that the post-measurement states of the last partite are still nonorthogonal. This situation leads to another set
   {\small	$$\mathcal{S}''=\{|\psi_1\rangle|\alpha''\rangle,|\psi_2\rangle|\alpha''\rangle,|\psi_3\rangle|\alpha''\rangle,|\psi_4\rangle|\beta''\rangle,|\psi_5\rangle|\beta''\rangle,
		|\psi_6\rangle|\beta''\rangle\}.$$ }
\noindent being distinguished. Noticing that the set $ \mathcal{S}''$   shares the same property as the original set $\mathcal{S}$.

The significant point is that the above  process cannot stop in any finite rounds of measurements. However, we are only allowed to perform  a finite rounds of measurements.  Therefore, we deduce that the given set $\mathcal{S}$ is indeed LOCC indistinguishable. \qed
	
\vskip 4pt

The above example shows  the main idea about how to construct a nonlocal set of product states. In fact,   we  can consider a much more general setting.

	\begin{theorem}\label{main}
		Let $\mathcal{S}_o=\{|\psi_i\rangle \ \big| 1\leq i\leq k\}$ be a set of  mutually orthogonal product states in $\mathbb{C}^{d_1}\otimes\mathbb{C}^{d_2}\otimes\cdots\otimes\mathbb{C}^{d_m}$. Suppose the set $\mathcal{S}_o$ is LOCC indistinguishable because all the local systems can not start with a nontrivial measurement to preserve the orthogonality of these given states. Let $|\alpha\rangle, |\beta\rangle$ be two nonorthogonal states in $\mathbb{C}^{d_{m+1}}$. Then the set of states $\mathcal{S}:=\{|\psi_1\rangle|\alpha_1\rangle,...,|\psi_k\rangle|\alpha_k\rangle\}$ is also LOCC indistinguishable as $m+1$ parties with $ |\alpha_1\rangle, \cdots, |\alpha_k\rangle\in \{|\alpha\rangle,|\beta\rangle\} $.
	\end{theorem}
	\noindent\emph{ Proof}:  By assumption, each of the first $m$ parties cannot be the first one to take a nontrivial measurement. Hence, the $(m+1)$-th system should be the first one to perform a nontrivial measurement $\mathcal{M}=\{M_i\}_{i=1}^n$.
	By lemma \ref{lemm}, there exists at least one $i$ such that
	$|\alpha'\rangle=M_i|\alpha\rangle, |\beta'\rangle= M_i|\beta\rangle$ are not orthogonal to each other. Normalizing the two states $|\alpha'\rangle, |\beta'\rangle$ to be $|\widehat{\alpha}'\rangle, |\widehat{\beta}'\rangle$, we redenote them to be $|\alpha'\rangle, |\beta'\rangle$.
	
	After the first round of measurement with the outcome $\text{``}i \text{ ''}$,  the post-measurement states are just
	$$\mathcal{S}'=\{|\psi_1\rangle|\alpha'_1\rangle,...,|\psi_k\rangle|\alpha'_k\rangle\}$$ with $|\alpha'_1\rangle, ...,|\alpha'_k\rangle\in \{|\alpha'\rangle, |\beta'\rangle\}$. In order to perfectly distinguish the states $\mathcal{S}$, we have to perfectly distinguish the states $ \mathcal{S}'$. It is obvious that $\mathcal{S}'$ shares the same properties with $\mathcal{S}$ including the following: the first $m$ local parts cannot start with   a nontrivial measurement, and the local states of the $(m+1)$-th system are just chosen from a set with two nonorthogonal states. With a  similar argument for the set $ \mathcal{S}'$,    there exists an outcome $\text{``}j \text{ ''}$ of $(m+1)$-th system measurement such that the post-measurement states of the last partite are still nonorthogonal. This situation leads to another set  $\mathcal{S}''$ being distinguished, which shares the same property as the original set $\mathcal{S}$. And this  process cannot stop in any finite rounds of measurements. However, we are only allowed to perform finite measurement rounds. To obtain a concise observation of the proof, we show a more intuitive figure (see Fig. \textcolor[rgb]{0.00,0.00,1.00}{1}). Hence, we deduce that the given set $\mathcal{S}$ is indeed LOCC indistinguishable. \qed

	\begin{figure}[h]
		\includegraphics[width=0.5\textwidth,height=0.37\textwidth]{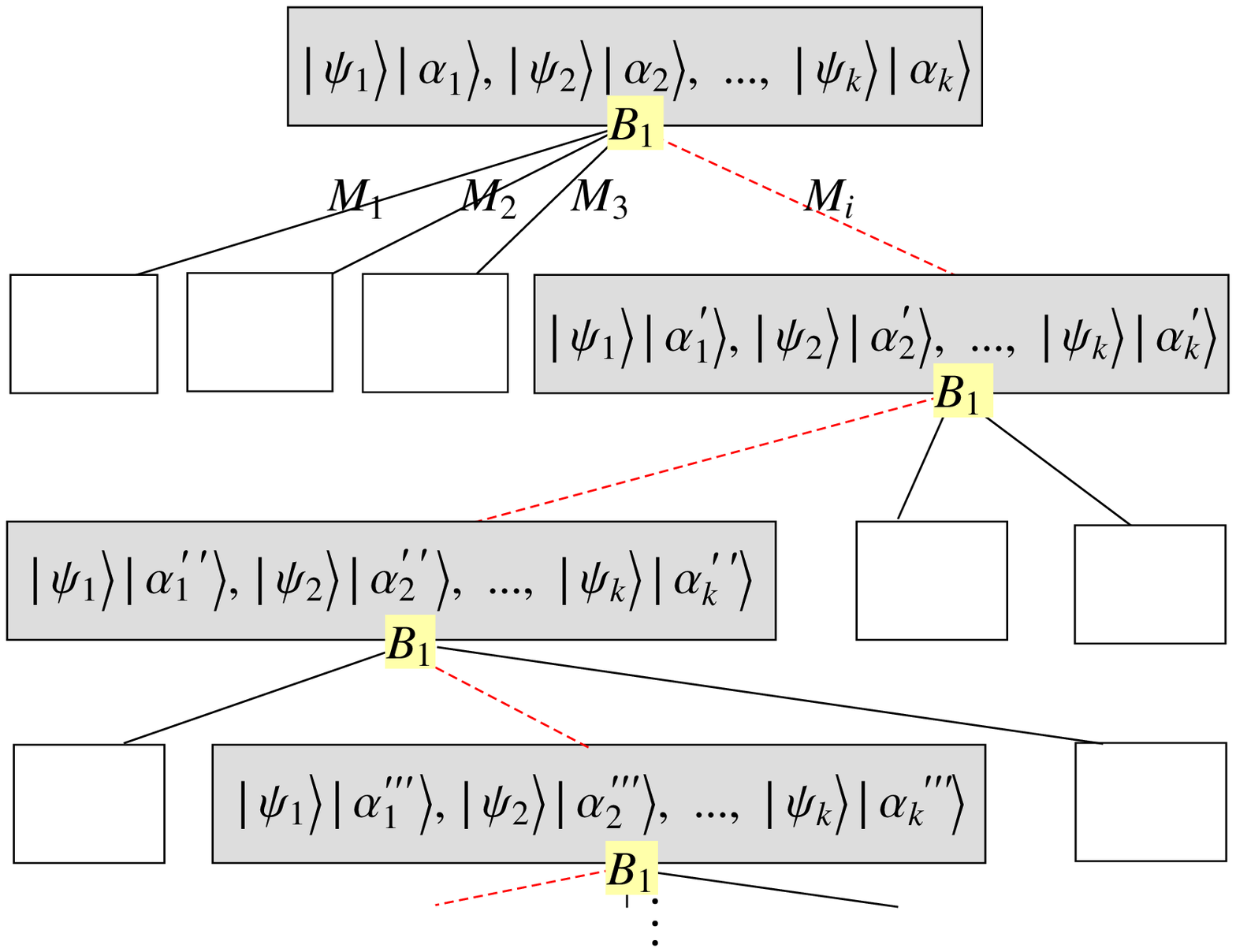}
\begin{adjustwidth}{4mm}{4mm}
\noindent{\textbf{Fig. 1:}{\small\emph{ This is a sketch map to show the distinguishing protocol of theorem \ref{main}. Here we use $A_1,A_2,...,A_m$ to denote the first $m$ parties and $B_1 $  to denote   the last partite. There exists an infinite sequence of outcomes (indicated by the red dotted line) performed by the last partite. Moreover, the post-measurement states along these outcomes are   pictured by the grey rectangle. }}}
\end{adjustwidth}	
	\end{figure}

	\vskip 5pt
Denote $|S|$ to be the cardinality of a set $S$.  The two sets $\mathcal{S}_o, \mathcal{S}$ in theorem \ref{main} satisfy $|\mathcal{S}_o|=|\mathcal{S}|$.
It is interesting to construct LOCC indistinguishable set with vary number of states.  Based on  the  current scheme, we give some results related to this problem.

We  maximally extent  $\{|\psi_1\rangle,...,|\psi_k\rangle\}$  by  orthogonal product states in $\mathbb{C}^{d_1}\otimes\mathbb{C}^{d_2}\otimes\cdots\otimes\mathbb{C}^{d_m}$ and denote such a target set to be  $\mathcal{S}'_o=\{|\psi_1\rangle,...,|\psi_k\rangle,|\psi_{k+1}\rangle..., |\psi_N\rangle\}$.  That is, the states in $\mathcal{S}'_o$ are mutually orthogonal product states and there is no nonzero product state  which is  orthogonal to all the states of $\mathcal{S}'_o$. If $N=k$, then $\mathcal{S}_o$ itself is an UPB (whose definition can be found in Ref. \cite{Ben99b}). If $N=\prod_{i=1}^m d_i$, then $\mathcal{S}_o$ is completable. Generally speaking,  $N$  lies between $k$ and $\prod_{i=1}^m d_i$. Let $\{|e_1\rangle, |e_2\rangle,\cdots,  |e_{d_{m+1}}\rangle\}$ be an orthonormal basis of $\mathbb{C}^{d_{m+1}} $ with $$|e_1\rangle=|\alpha\rangle, \text{  and  } \text{span}_\mathbb{C}\{|e_1\rangle, |e_2\rangle\}=\text{span}_\mathbb{C}\{|\alpha\rangle, |\beta\rangle\}.$$ Denote $|\alpha^\perp\rangle, |\beta^\perp\rangle$ to be the unique states (up to a phase) in  $\text{span}_\mathbb{C}\{|\alpha\rangle, |\beta\rangle\}$  which is orthogonal to $|\alpha\rangle, |\beta\rangle$, respectively. Then  the following set $\mathcal{S}_{ext}$  of product states in $\mathbb{C}^{d_1}\otimes\mathbb{C}^{d_2}\otimes\cdots\otimes\mathbb{C}^{d_m}\otimes\mathbb{C}^{d_{m+1}}$
	$$\begin{array}{l}
|\psi_1\rangle|\alpha_1\rangle,...,|\psi_k\rangle|\alpha_k\rangle,|\psi_{k+1}\rangle|\alpha\rangle,..., |\psi_N\rangle|\alpha\rangle,  \\
|\psi_1\rangle|\alpha_1^\perp\rangle,...,|\psi_k\rangle|\alpha_k^\perp\rangle,|\psi_{k+1}\rangle|\alpha^\perp\rangle,..., |\psi_N\rangle|\alpha^\perp\rangle, \\
|\psi_1\rangle|e_3\rangle,...,|\psi_k\rangle|e_3\rangle,|\psi_{k+1}\rangle|e_3\rangle,..., |\psi_N\rangle|e_3\rangle,  \\
	\ \ \ \ \ \ \ \ \ \ \ \ \ \ \ \ \ \  \vdots\\
|\psi_1\rangle|e_{d_{m+1}}\rangle,...,|\psi_k\rangle|e_{d_{m+1}}\rangle,|\psi_{k+1}\rangle|e_{d_{m+1}}\rangle,..., |\psi_N\rangle|e_{d_{m+1}}\rangle \end{array}
	$$
	are pairwise orthogonal  to each other.

It is clearly that adding more states into a set increases
the difficulty to discriminate them.
Therefore, any set $\widetilde{\mathcal{S}}$ with $\mathcal{S} \subseteq\widetilde{\mathcal{S}}\subseteq \mathcal{S}_{ext}$ are also locally indistinguishable by LOCC.
Notice that $|	\widetilde{\mathcal{S}}|$ could be any integer lie between $k$  and $Nd_{m+1}$. If $\mathcal{S}$ is completable, then $N$ can be chosen to be $\prod_{i=1}^{m}d_i$. Hence,
$|	\widetilde{\mathcal{S}}|$ could be any integer lie between $k$  and $\prod_{i=1}^{m+1} d_i$.

	From the above argument, it is not difficult to show that if the original set $\{|\psi_1\rangle,...,|\psi_k\rangle\}$ is completable,  then  $\{|\psi_1\rangle|\alpha_1\rangle,...,|\psi_k\rangle|\alpha_k\rangle\}$ is also completable.   For this setting(completable case), this  approach of  provides a large set of separable measurements, which cannot be accomplished by LOCC measurement.

	\begin{theorem}\label{main2}
		Let $\mathcal{S}_o=\{|\psi_i\rangle \ \big| 1\leq i\leq k\}$ be a set of mutually orthogonal product states in $\mathbb{C}^{d_1}\otimes\mathbb{C}^{d_2}\otimes\cdots\otimes\mathbb{C}^{d_m}$. Suppose the set $\mathcal{S}_o$ is LOCC indistinguishable because all the local systems cannot start with a nontrivial measurement in order to preserve the orthogonality of these given states. Let $|\alpha^{(j)}\rangle, |\beta^{(j)}\rangle$ be  two nonorthogonal  states in  $\mathbb{C}^{d'_j}$ for $j=1,2,...,n$. Then the set $\mathcal{S}$ of states $$\{|\psi_1\rangle\otimes(\otimes_{j=1}^n|\alpha^{(j)}_1\rangle),...,|\psi_k\rangle\otimes(\otimes_{j=1}^n|\alpha^{(j)}_k\rangle)\}$$ in $ (\otimes_{i=1}^m\mathbb{C}^{d_i})\otimes(\otimes_{j=1}^n\mathbb{C}^{d'_j})$
		 is also LOCC indistinguishable as $m+n$ parties where $ |\alpha^{(j)}_1\rangle, \cdots, |\alpha^{(j)}_k\rangle\in \{|\alpha^{(j)}\rangle,|\beta^{(j)}\rangle\} $ for $j=1,2,...,n$.
	\end{theorem}
	\noindent\emph{ Proof}:  The idea of the proof is similar with that of theorem \ref{main}.  We call a set $S_0$ of mutually orthogonal product states in $(\otimes_{i=1}^m\mathbb{C}^{d_i})\otimes(\otimes_{j=1}^n\mathbb{C}^{d'_j})$  shares the property $\mathcal{P}(m,n)$ if it satisfies the following two conditions:
	\begin{enumerate}[(i)]
		\item To preserve the orthogonality of $S_0$, each of the first $m$ parties could only do a trivial measurement.
		\item For any state $S_0$, its local states of the last $n$ parties are just chosen from two pure nonorthogonal states of the corresponding system.
	\end{enumerate}
	
	By assumption, the set $\mathcal{S}$ shares the property $\mathcal{P}(m,n)$. Thus, each of  the first $m$ parties cannot be the first one to take a nontrivial measurement. Hence, one of the last $n$ systems should be the first to take the nontrivial measurement. However, by lemma \ref{lemm}, for any such measurement, there is an outcome such that the post-measurement states $\mathcal{S}'$ also share property $\mathcal{P}(m,n)$. To perfectly distinguish the states $\mathcal{S}$, we have to perfectly distinguish those states $\mathcal{S}'$.
	
	With the same argument, we could show that in order to perfectly distinguish the states $\mathcal{S}'$,  we have to perfectly distinguish another set $\mathcal{S}''$, which also shares property $\mathcal{P}(m,n)$. The significance is that this process cannot stop in any finite rounds of measurements.
	However, we are only allowed to play a finite rounds of measurements by the definition of LOCC. Hence, we deduce that the given set $\mathcal{S}$ is LOCC indistinguishable. Fig. \textcolor[rgb]{0.00,0.00,1.00}{2} is a more intuitive figure to show the protocol presented in the proof.
	
\qed

\begin{figure}[h]\centering
		\includegraphics[width=0.49\textwidth,height=0.40\textwidth]{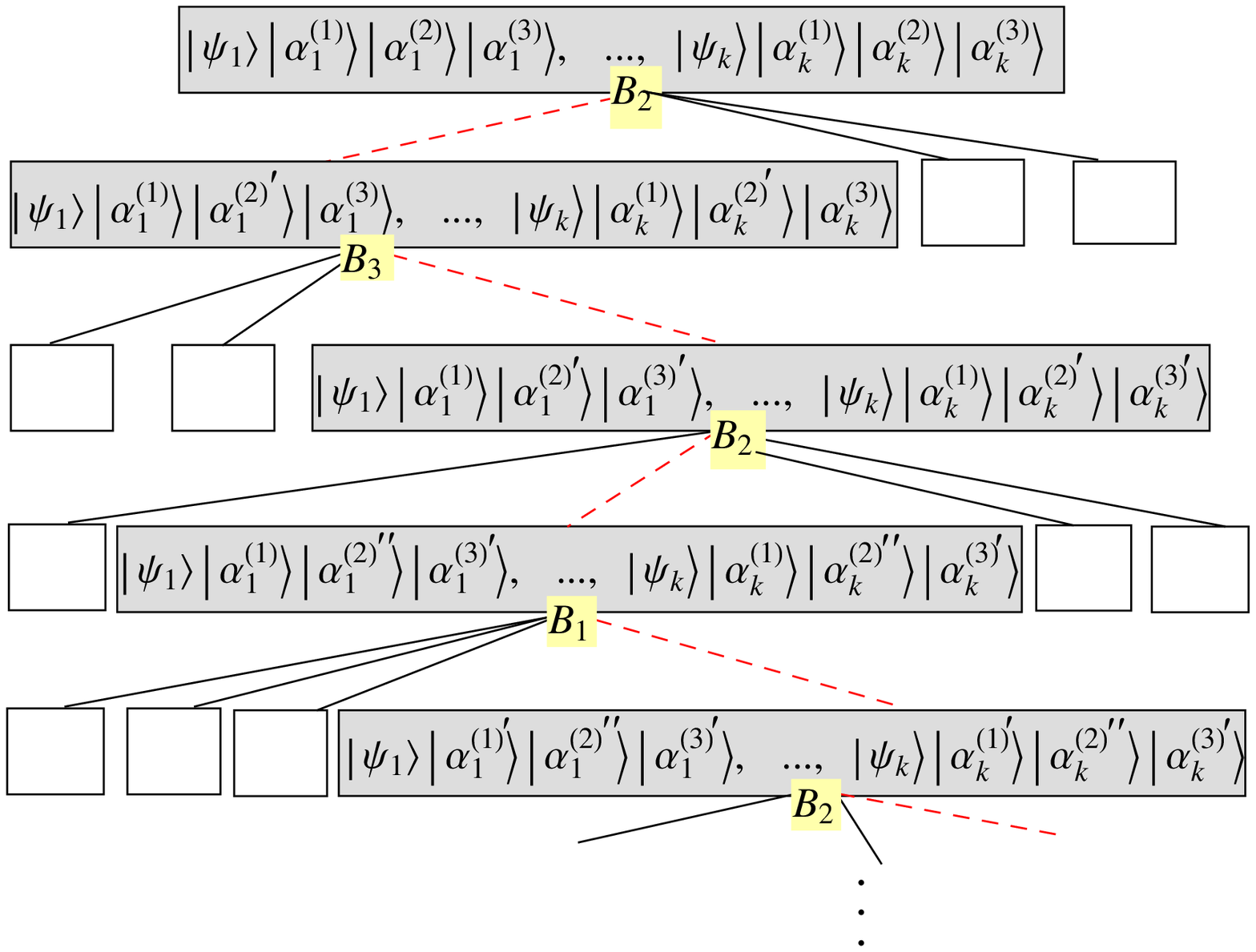}
\begin{adjustwidth}{4.4mm}{4.4mm}
\noindent{\textbf{Fig. 2:} {\small\emph{This is a sketch map to show the distinguishing protocol of theorem \ref{main2} with $n=3$. Here we use $A_1,$ $A_2,...,$ $A_m$ to denote the first $m$ parties and $B_1, B_2,B_3 $ to denote  the last three parties. The significant point is   that there exists an infinite sequence of outcomes (indicated by the red dotted line) performed by the last three parties. And these parties send a set with the $\mathcal{P}(m,3)$ property to another set with the same property.  Moreover, the post-measurement states along these outcomes are   pictured by the grey rectangle.} }}
\end{adjustwidth}
	\end{figure}

	\vskip 5pt
	
	\begin{corollary}\label{cor2}
		Suppose $\{|\psi_1\rangle,...,|\psi_k\rangle\}\subseteq \mathbb{C}^{d_1}\otimes\mathbb{C}^{d_2}\otimes\cdots\otimes\mathbb{C}^{d_m}$ is LOCC indistinguishable because no local system could play a nontrivial measurement in order to preserve the orthogonality of the given states. Let $|\alpha^{(j)}\rangle $ be any states in $\mathbb{C}^{d'_{j}}$ for $j=1,2,...,n$, then the set of states $\{|\psi_1\rangle\otimes_{j=1}^n |\alpha^{(j)}\rangle,...,|\psi_k\rangle\otimes_{j=1}^n |\alpha^{(j)}\rangle\}$ is also LOCC indistinguishable as $m+n$ parties.
	\end{corollary}

To better compare the result in corollary \ref{cor2}  with those in  \cite{Zhang17}, here we   quote one of the main results from  reference  \cite{Zhang17}.	
``Suppose $\{|\varphi\rangle_i=|x\rangle_i|y\rangle_i, i=1,2,...,l\}$ is a set of orthogonal
product states in $d\otimes d$, which cannot be distinguished
by LOCC because two parties cannot start a nontrivial measurement.
 Then, we give the following multipartite orthogonal product states
 in $d_1\otimes d_2\otimes\cdots \otimes d_n$ $(d_{1,2,...,n}=d\geq 3,
 n\geq 4).$
$$
   \begin{array}{l}
     |\varphi_i\rangle=|x\rangle_i|y\rangle_i|b\rangle_3\cdots |b\rangle_{n-1}
     |a\rangle_n, \\
          |\varphi_{i+l}\rangle=|x\rangle_i|a\rangle_2|y\rangle_i\cdots |b\rangle_{n-1}
     |b\rangle_n,\\
     \ \ \ \ \ \ \  \vdots \\
          |\varphi_{i+ml}\rangle=|x\rangle_i|b\rangle_2 \cdots|a\rangle_{m+1}|y\rangle_i
          \cdots      |b\rangle_n,\\
     \ \ \ \ \ \ \   \vdots \\
         |\varphi_{i+(n-2)l}\rangle=|x\rangle_i|b\rangle_2 \cdots|b\rangle_{n-2}|a\rangle_{n-1}
             |y\rangle_i
   \end{array}
$$
where $i=1,2,\cdots, l,\ 2\leq m\leq n-3, \ |a\rangle_j$ is orthogonal to $|b\rangle_j, (2\leq j\leq n)$ in $d$-dimension quantum system."
Then the theorem 1 of Ref. \cite{Zhang17} asserts that the above $(n-1)l$ states cannot be perfectly distinguished by LOCC.

Fixed $m\in\{0,...,n-2\}$, we can obtain that the set $\mathcal{S}_m$
$$\{|\varphi_{i+ml}\rangle \ \big | \   i=1,2,...,l\}$$
          cannot  be distinguished by LOCC from corollary 1. Therefore, the result of the corollary improves the results of study by Zhang \cite{Zhang17} from the following two aspects.
	\begin{enumerate}[(a)]
		\item All the unit states $|\alpha^{(j)}\rangle\in\mathbb{C}^{d'_{j}}$ are chosen  arbitrarily and independently of each other.
		\item  The set corresponding to the above  corollary is only a subset of those in \cite{Zhang17}.
	\end{enumerate}
In \cite{Zhang17}, Zhang \emph{et al.} showed that to preserve the orthogonality of states in $\bigcup_{m=0}^{n-2} \mathcal{S}_m$ any partite could only do a trivial measurement.
Then they can obtain their statement. However, 	 to   preserve the orthogonality of states in $\mathcal{S}_m$   there are still some parties  which could start with nontrivial measurement. Hence their methods fail to deal with the local distinguishability of $\mathcal{S}_m$.
	\vskip 5pt
It is natural to ask whether   the results above can be
generalized to cases with three or more nonorthogonal states.
A  ``simply generalized" version of theorem \ref{main} like replacing ``Let $|\alpha\rangle, |\beta\rangle$  be two $\cdots \cdots$  $ |\alpha_1\rangle, \cdots, |\alpha_k\rangle\in \{|\alpha\rangle,|\beta\rangle\} $"  in theorem \ref{main} by ``Let $|\alpha\rangle, |\beta\rangle, | \gamma\rangle$  be three $\cdots \cdots$  $ |\alpha_1\rangle, \cdots, |\alpha_k\rangle\in \{|\alpha\rangle,|\beta\rangle, |\gamma\rangle\} $" may not be true, an example of which is provided below.
	
	\vskip 2pt
	
	\begin{example}\label{exa} With the same notation in equation (\ref{states}), and
		let $|\alpha\rangle=|0\rangle,|\beta\rangle=\frac{1}{2}|0\rangle+\frac{\sqrt{3}}{2}|1\rangle,|\gamma\rangle=\frac{1}{2}|0\rangle-\frac{\sqrt{3}}{2}|1\rangle$.
		Then the set of states $$|\psi_1\rangle|\alpha\rangle,|\psi_2\rangle|\alpha\rangle,|\psi_3\rangle|\beta\rangle,|\psi_4\rangle|\beta\rangle,|\psi_5\rangle|\gamma\rangle,
		|\psi_6\rangle|\gamma\rangle$$
		is LOCC distinguishable.
	\end{example}
	\noindent\emph{ Proof:} Firstly, the fourth partite performs  the  measurement $\{ M_1, M_2\}$   with
	$$M_1^\dagger M_1=\left[
	\begin{array}{cc}
	\frac{1}{2} & \frac{1}{2\sqrt{3}} \\[1mm]
	\frac{1}{2\sqrt{3}} & \frac{1}{2} \\
	\end{array}
	\right],
	M_2^\dagger M_2=\left[
	\begin{array}{cc}
	\frac{1}{2} & \frac{-1}{2\sqrt{3}} \\[1mm]
	\frac{-1}{2\sqrt{3}} & \frac{1}{2} \\
	\end{array}
	\right].
	$$
	Clearly,
	$M_1^\dagger M_1+M_2^\dagger M_2=I_2 $ and both $M_1^\dagger M_1$ and $M_2^\dagger M_2$ are semipositive definite matrices.  Moreover, it can be easily verified that
	$$\langle \alpha| M_1^\dagger M_1| \gamma\rangle=0,\ \ \  \langle \alpha| M_2^\dagger M_2| \beta\rangle=0.$$
	According to these two equations, we have the following protocol to distinguish the six states in the example.
	\begin{itemize}
		\item If the measurement outcome is $\text{``}1\text{''}$, Danny (the fourth system) continuously performs her measurement according to the basis $\{ M_1| \alpha\rangle,M_1| \gamma\rangle\}$ (must be normalized). Moreover, the outcome $\text{``}\gamma\text{''}$ could eliminate the states  $|\psi_1\rangle|\alpha\rangle,|\psi_2\rangle|\alpha\rangle$, and the first three parties could distinguish the other four. The outcome $\text{``}\alpha\text{''}$ could eliminate the states $|\psi_5\rangle|\gamma\rangle,|\psi_6\rangle|\gamma\rangle$; then, the first three parties could distinguish the remaining four states.
		\item If the measurement outcome is $\text{``}2\text{''}$, Danny continuously measures according to the basis $\{ M_2| \alpha\rangle,M_2| \beta\rangle\}$. The outcome $\text{``}\beta\text{''}$ could eliminate the states $|\psi_1\rangle|\alpha\rangle,|\psi_2\rangle|\alpha\rangle$, and the first three parties could distinguish the other four. The outcome  $\text{``}\alpha\text{''}$ could eliminate the states $|\psi_5\rangle|\beta\rangle,|\psi_6\rangle|\beta\rangle$, then the first three parties could distinguish the remaining four states.
	\end{itemize}	\qed
	
\noindent {\textbf{Remark:}} It is much of interest to ask whether  it is  possible to generate new
LOCC indistinguishable set by appending the system with  three or more
``appropriately selected"  nonorthogonal states. Up to now, we haven't been able to solve this problem.
	
	\section{conclusion and discussion}\label{for}
	In this paper, we present a new method for deriving sets of  locally indistinguishable product states in multipartite quantum systems. The method is based on the following observation: given two nonorthogonal states; there exists at least one outcome of any measurement such that the post-measurement states corresponding to the two states are still nonorthogonal. Given any set of LOCC indistinguishable product states because all the local systems can not start with a nontrivial measurement, we can extend it  by another partite with the local state being chosen from a set with two nonorthogonal states. However, with an additional partite joined in, the set is still LOCC indistinguishable.  Based on this scheme, we  show that it is possible to construct LOCC indistinguishable set with vary number of states. After that we generalize this with more parties joined in and show the   target set is still LOCC indistinguishable.   It can be seen that   some parties can perform some nontrivial measurements.  Hence, the sets are completely different from those constructed before.
	This result widen our knowledge of nonlocality without entanglement to some extent.
	
\vskip 5pt

	Moreover, we show that if the original given set is completable, the set with more parties is also completable. The local indistinguishability implies that these sets could be identified perfectly by separable operators. All these give nonlocal separable operations, that is, separable operations that cannot be implemented by LOCC.

\vskip 5pt

There are also some interesting questions remain unsolved. For example, is it possible to generate new
LOCC indistinguishable set by appending the system with  three or more ``appropriately selected" nonorthogonal states?
If so, this would   increase our freedom to construct set of  locally indistinguishable product states.
\vskip 5pt

	\vspace{2.5ex}
	
\noindent{\bf Acknowledgments}\, \, The authors thank the  referee  for asking several interesting questions to improve our previous  version of this  paper.
The work is supported by the  NSFC 11571119.

\vspace{2.5ex}
	
	{
		$${\text{\textbf{APPENDIX A}}}$$
	}

	\noindent\emph{ Proof of lemma \ref{lemm}}: Since $\langle \alpha|\beta\rangle\neq 0$, we can write $|\beta\rangle$ as the form
	$$|\beta\rangle=\lambda |\alpha\rangle+\delta|\alpha^\perp\rangle, \ \  \lambda \neq 0 $$ where $|\alpha^\perp\rangle \in \mathbb{C}^{d } $ is a unit vector and $\langle \alpha|\alpha^\perp\rangle= 0.$   Suppose $\{|\alpha_1\rangle, |\alpha_2\rangle,\cdots,  |\alpha_d \rangle\}$ are  orthonormal basis of $\mathbb{C}^{d } $ and $|\alpha_1\rangle=|\alpha\rangle,|\alpha_2\rangle=|\alpha^\perp\rangle.$ Under   this given basis, the matrix $ M_i^\dagger M_i$
	can be written as the form
	$$
	\left[
	\begin{array}{cccc}
	m^{(i)}_{11} &  m^{(i)}_{12} &    \cdots &  m^{(i)}_{1d} \\[2mm]
	m^{(i)}_{21} &  m^{(i)}_{22} &   \cdots &  m^{(i)}_{2d} \\[1mm]
	\vdots  & \vdots &    \ddots & \vdots \\
	m^{(i)}_{d1} &  m^{(i)}_{d2} &   \cdots &  m^{(i)}_{dd} \\
	\end{array}
	\right].
	$$
	Suppose the claim of our lemma is not true. That is, for each $i$, at least one of the following three elements is zero:
	$$
	\langle \alpha|  M_i^\dagger M_i|\alpha\rangle,\langle \alpha|  M_i^\dagger M_i|\beta\rangle,\langle \beta|M_i^\dagger   M_i|\beta\rangle.$$
	Hence
	the matrices $\{ M_i^\dagger M_i\}_{i=1}^n$ can be separated into three classes as follows:
	\begin{enumerate}[(i)]
		\item $\langle \alpha| M_i^\dagger M_i|\alpha\rangle=0$ implies that $ m^{(i)}_{11}= m^{(i)}_{12}=0$.
		\item  $\langle \alpha| M_i^\dagger M_i|\beta\rangle=0$ implies that $\lambda  m^{(i)}_{11}+\delta  m^{(i)}_{12}=0.$
		\item  $\langle \beta| M_i^\dagger M_i|\beta\rangle=0$ implies that $M_i|\beta\rangle=\textbf{0}.$ Hence $M_i^\dagger M_i|\beta\rangle=\textbf{0}$.  This also implies
		the relation between $ m^{(i)}_{11}$ and $ m^{(i)}_{12}$: $\lambda  m^{(i)}_{11}+\delta  m^{(i)}_{12}=0.$
	\end{enumerate}
	
	Hence, we conclude that the three cases could give the same relation: $\lambda  m^{(i)}_{11}+\delta  m^{(i)}_{12}=0.$ Since this holds for all $i\in \{1,2,...,n\}$, we have
	\begin{equation}\label{relation}
	\lambda \sum_{i=1}^n m^{(i)}_{11}+\delta \sum_{i=1}^n m^{(i)}_{12}=0.
	\end{equation}
	On the other hand, the measurement definition naturally  gives
	$$\sum_{i=1}^n M_i^\dagger M_i=I_d.
	$$
	Particularly, by considering the entrywise equalities, we obtain
	\begin{equation}\label{relation2}
	\sum_{i=1}^n m^{(i)}_{11}=1, \  \sum_{i=1}^n m^{(i)}_{12}=0.
	\end{equation}
	Then, equations (\ref{relation}) and (\ref{relation2}) contradict each other as $\lambda\neq 0$.
	Hence, this completes the proof claimed in the lemma.
	
\qed

\end{document}